# Liquid Metals Routes towards Making Superconductors

Chen Hua[a,b], Wendi Bao[a,b], Minghui Guo[a], Jing Liu[a,b*]

[a]*State Key Laboratory of Cryogenic Science and Technology and Beijing Key Laboratory of Cryobiomedicine, Technical Institute of Physics and Chemistry, Chinese Academy of Sciences, Beijing 100190, China.*

[b]*School of Future Technology, University of Chinese Academy of Sciences, Beijing 100049, China.*

*Correspondence: jliu@mail.ipc.ac.cn

## Abstract

We conceive liquid-metal-derived superconductors (LMDS) as a unified paradigm that enables the quick fabrication of superconducting materials under near-ambient conditions through introducing room-temperature liquid metals (LMs) as dynamic metallic reaction media. In this framework, LMs serve as solvents, dopant reservoirs, interfacial mediators, and structural templates that lower the barrier to forming superconducting-relevant material states. This paradigm integrates LM-enabled pathways for producing bulk alloys, printed films, two-dimensional confined phases, interconnect geometries, and nanodroplets. Their liquid-state processability enables near-room-temperature patterning, reconfiguration, and compositional control, whereas superconducting functionality is established in cooled LM-derived states such as solidified alloys, doped films, amorphous/glassy phases, nanoconfined structures, and interfacially reconstructed layers. We further outline a data-driven LM materials genome that unifies composition, structure, ground-state quantities, interaction parameters, and macroscopic properties to accelerate predictive modeling and inverse design of LMDS. Beyond processing advantages, LMs provide an experimental platform for examining superconductivity in amorphous, nanoconfined, and dynamically disordered states and for revisiting the longstanding question of whether true superconductivity can exist in liquid state. This perspective positions LMs as a fertile and energy-efficient route toward reconfigurable and potentially transformative superconducting technologies.

## 1 Introduction

Superconductors are vital for energy, medicine, and quantum technology but often limited by complex fabrication, fragility, and cryogenic cost. Over recent decades, significant progress has been achieved for both ceramic and metallic superconductors such as $YBa_2Cu_3O_{7-x}$ (YBCO), NbTi, and $Nb_3Sn$, yielding materials with excellent critical current performance, stability, and scalability. However, so far most fabrication routes remain energy intensive and technically demanding.

Bulk synthesis methods, such as solid-state reaction, high-pressure synthesis[1], sol-gel[2], co-precipitation, and flux[3], typically require temperatures above 1000 K, long reaction times, and precise thermal control. Thin film techniques[4] such as pulsed laser deposition, electron beam evaporation, molecular beam epitaxy, metal-organic chemical vapor deposition, chemical vapor deposition[5], and magnetron sputtering offer atomic scale precision but demand high temperatures, ultrahigh vacuum, and lattice matched substrates. Exfoliation routes, including mechanical and liquid-phase approaches, face limitations in scalability and uniformity despite their simplicity.

Powder and fiber methods such as electrospinning[6], solution blow spinning[7], and freeze-drying enable scalable structures but still require sintering near 1200 K and often weaken mechanical integrity. Even rapid routes such as combustion and aerosol[8] processes involve localized high temperature reactions.

Common challenges including high temperature and pressure requirements, lengthy processing, and mechanical fragility arise from the brittle ceramics and high melting alloys that dominate existing superconductors, motivating the development of simpler, energy efficient fabrication strategies.

LMs combine metallic bonding and electrical conductivity with exceptionally low melting points ($\boldsymbol{T_m}$) and high atomic mobility, placing them in a distinctive regime characterized by soft lattices, strong structural disorder, and dynamic atomic rearrangements. We focus on how these features delineate a physically motivated design space for superconductors, and how data-driven approaches can be used to organize existing experimental observations and theoretical calculations and to guide future exploration. Beyond processing considerations, LMs provide a well-defined testbed for probing the structural origins of superconductivity and its persistence across crystalline, amorphous, nanoconfined, and liquid-like states. On this basis, we introduce the concept of LMDS as a broad theoretical framework encompassing both superconductors fabricated or engineered through liquid-metal-assisted routes and superconducting states directly arising in LMs themselves. In the former case, LMs serve as solvents, dopants, interfacial mediators, or reactive environments for superconducting material synthesis and engineering. In the latter case, LMDS includes superconducting systems associated with LMs, such as amorphous or nanoconfined phases.

## 2 Clues to LMDS

The discovery of superconductivity provides an instructive historical clue. In 1911, Kamerlingh Onnes observed zero electrical resistance in mercury (a typical room-temperature LM) at 4.2 K, marking the first confirmation of superconductivity.

Subsequent efforts sought materials with higher superconducting transition temperature ($\boldsymbol{T_c}$) and easier fabrication to expand applications and reduce costs. High-throughput and machine-learning studies based on chemical formula descriptors identified the $\boldsymbol{T_m}$ as the most important feature for predicting $\boldsymbol{T_c}$ [9]. This correlation is thermodynamically consistent: Both solid–liquid and normal–superconducting transitions reflect competing order and disorder but differ in their ordering nature. Melting is a first-order transition with abrupt changes in volume, enthalpy and shear modulus, whereas superconductivity appears continuously through quantum coherence.

To examine the relationship between $\boldsymbol{T_c}$ and $\boldsymbol{T_m}$, selected properties of all metallic elements were analyzed, with elemental data obtained from ElementData[10], a curated database compiled from diverse sources and enhanced by Wolfram Research. Due to the skewed data distribution, Spearman correlation coefficients were calculated and shown in the upper-right panel of **Fig. 1**. This non-parametric measure, based on data ranks rather than raw values, quantifies the strength and direction of monotonic relationships between variables and is robust to outliers. The Spearman correlation coefficient $r_s$ is calculated as:

$$r_s = 1 - \frac{6\sum d_i^2}{n(n^2-1)} \tag{1}$$

$$d_i = R(x_i) - R(y_i) \tag{2}$$

$$R(x_i) = \begin{cases} Position(x_i\ in\ S), & \text{if only one } x_i \\ \dfrac{\sum_{j \in P(x_i)} j}{|P(x_i)|}, & \text{if multiple } x_i \end{cases} \tag{3}$$

where $n$ is the number of samples, $S = \{s_1, s_2, \dots, s_n\}$ is the ascending ordering of the set of variables $X = \{x_1, x_2, \dots, x_n\}$, $P(x_i) = \{j | s_j = x_i\}$ is the set of positions of the value $x_i$ in the sorted sequence S, and $|P(x_i)|$ is the number of elements in the set $P(x_i)$.

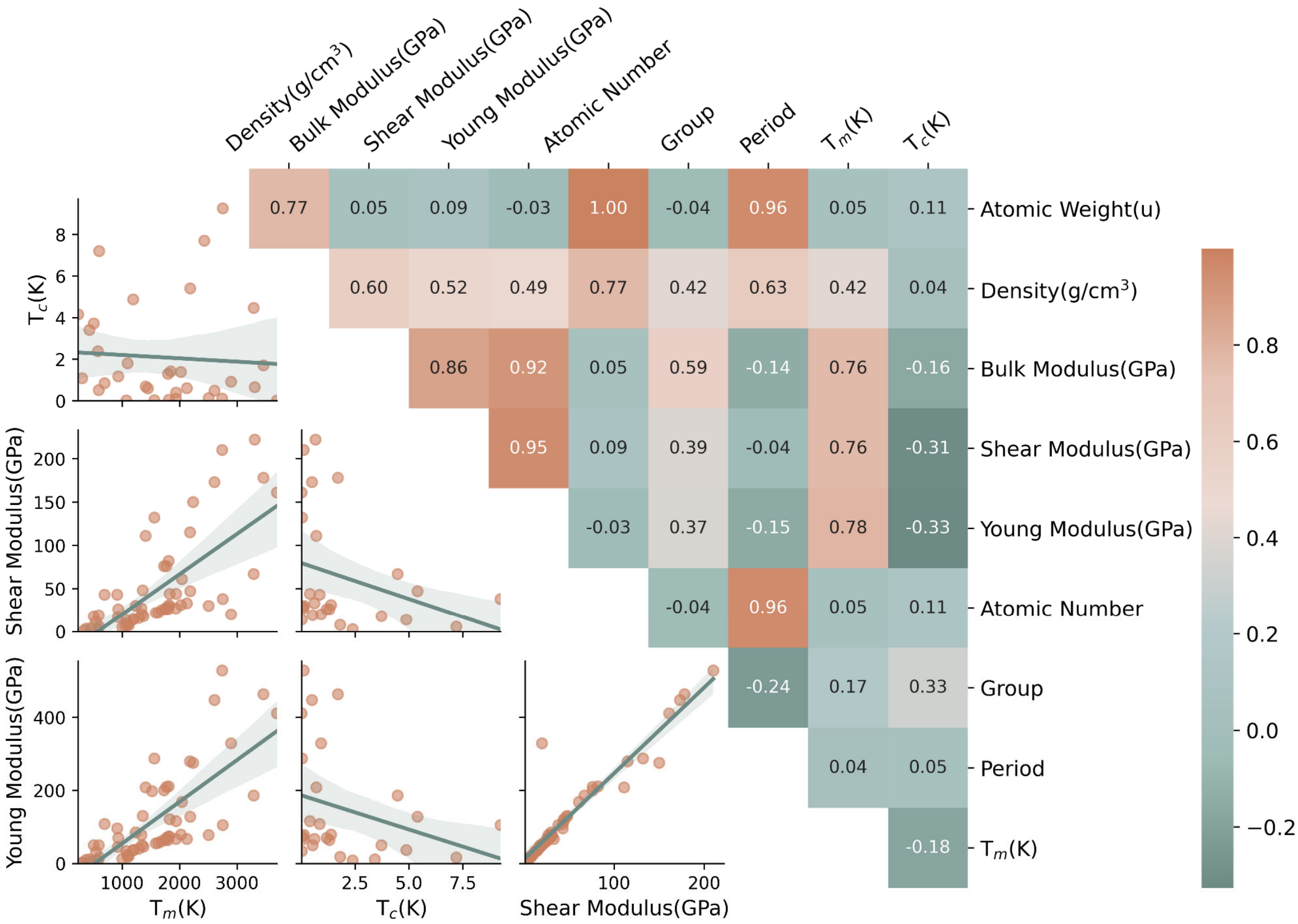

**Fig. 1 Property correlation atlas for all metal elements.**

By analyzing the Spearman correlation coefficients, the shear and Young's moduli were found to be negatively correlated with $\boldsymbol{T_c}$ (–0.31, –0.33), indicating that softer metals tend to exhibit higher $\boldsymbol{T_c}$, consistent with the fluid-like nature of low-modulus systems. Both moduli show strong positive correlations with $\boldsymbol{T_m}$ (0.76, 0.78), suggesting that low rigidity corresponds to a lower $T_m$. Taken together, these trends imply a weak overall negative correlation between $T_m$ and $T_c$ (–0.18), inspiring that softer, lower-$T_m$ materials may provide a favorable environment for conventional superconductivity. The relationships among $T_c$, $T_m$, shear modulus, and Young's modulus were fitted using ordinary least squares, and the resulting linear fits of the variables are shown in the lower-left panel of **Fig. 1**, highlighting their positive and negative correlations.

The above trend can be qualitatively understood within the conventional phonon-mediated superconductivity framework derived from BCS theory[11], particularly through the Eliashberg–McMillan formalism[12]. In general, elemental metals with lower $T_m$ often tend to possess softer lattices and weaker interatomic bonding. According to McMillan theory, the electron–phonon coupling constant is given by

$$\lambda = \frac{N(0)\langle I^2 \rangle}{M\langle \omega^2 \rangle} \tag{4}$$

where $N(0)$ is the electronic density of states at the Fermi level, $\langle I^2 \rangle$ is the electron–phonon interaction matrix element, $M$ is the atomic mass, and $\langle \omega^2 \rangle$ is the averaged squared phonon frequency. Softer lattices generally correspond to lower phonon frequencies, thereby enhancing $\lambda$ through the inverse dependence on $\langle \omega^2 \rangle$. At the same time, lower phonon frequencies also reduce the Debye temperature according to

$$\Theta_D = \frac{\hbar \omega_D}{k_B} \tag{5}$$

where $\hbar$ is the reduced Planck's constant, $\omega_D$ is the Debye frequency, and $k_B$ is the Boltzmann constant. According to the McMillan equation,

$$T_c^{McMillan} = \frac{\Theta_D}{1.45} \exp\left[-\frac{1.04(1+\lambda)}{\lambda - \mu^*(1+0.62\lambda)}\right] \tag{6}$$

although lower phonon frequencies reduce the Debye temperature, the enhancement of electron–phonon coupling enters exponentially in the $T_c$ expression and may therefore partially compensate for the reduced Debye temperature, contributing to the empirical tendency observed in certain elemental metals.

The observed correlation between lower $T_m$, softer lattices, and superconducting tendencies applies mainly to conventional phonon-mediated superconductors within the BCS/Eliashberg–McMillan framework, rather than unconventional systems such as cuprates and Fe-based superconductors. Although this empirical correlation is derived primarily from elemental conventional superconductors, it provides a useful physical motivation for exploring LMDS, where low $\boldsymbol{T_m}$ and soft lattices may influence phonon-mediated superconducting behavior. We also acknowledge that major advances in modern superconductivity research have been driven by unconventional pairing mechanisms, strongly correlated electronic states, and intertwined quantum orders, which are beyond the scope of this perspective. Despite extensive progress across diverse superconductor families, LMDS remain largely unexplored. Although elemental metals at ambient pressure typically exhibit relatively low $\boldsymbol{T_c}$ values (mostly around 2 K and generally below 10 K, as shown in **Fig. 1**), alloying can substantially reduce $\boldsymbol{T_m}$, making LMDS an experimentally accessible platform for investigating possible superconducting behavior under chemically tunable conditions.

LMs are metallic fluids that remain stable and conductive near room temperature. Typical systems such as gallium, indium, and their eutectic alloys (EGaIn, Galinstan) combine metallic bonding with exceptional fluidity. More LM types such as bismuth based alloys (BiInSn, BiInSnZn) are also possible candidates. Their low $\boldsymbol{T_m}$ and soft lattices make them natural extensions of the metallic superconducting family, while their fluid state enables new modes of materials design and integration.

Superconductivity research involving LMs has developed from low-$\boldsymbol{T_m}$ metals to alloy, interface, and device systems. Early studies showed that amorphous Ga reaches a $\boldsymbol{T_c}$ of about 8.4 K, much higher than crystalline α-Ga, while amorphous Bi becomes superconducting near 6 K although crystalline Bi is essentially non-superconducting at ambient pressure[13,14]. These observations suggest that superconductivity can persist or even be enhanced in structurally disordered or metastable phases, motivating interest in LM-derived superconducting systems[15,16].

Recent studies have further expanded LM-related superconductivity. Ga-In-Sn nanodroplets, In-Sn alloys, and Ga-In-Sn or In-Bi-Sn solder systems show superconducting transitions around 6 K after cooling, composition tuning, or intermetallic formation[17–21]. Ga-containing compounds and confined/interfacial systems, including $Mo_4Ga_{20}As$, Ga-hyperdoped Ge, Ga-intercalated graphene, and graphene/trilayer-Ga/SiC heterostructures, demonstrate that Ga can act as a superconducting host, dopant, or interfacial mediator[22–26]. In particular, confined trilayer Ga in a graphene/SiC heterostructure exhibits orbital-hybridization-induced Ising-type superconductivity with $T_c \approx 4$ K and an in-plane upper critical field far beyond the Pauli limit[26]. In parallel, LM inks and printing approaches have enabled superconducting circuits and resonators fabricated under near-ambient conditions, although superconducting operation remains cryogenic[27,28]. Collectively, these studies illustrate the emerging and multidisciplinary nature of LM-related superconductivity research across materials synthesis, interface engineering, and device fabrication.

Building on these existing studies, **Fig. 2** summarizes LMs as adaptive metallic matrices that define a multifunctional design space for LMDS. The central disk organizes four roles: Dopant reservoir, solvent, interface mediator, and structural host/template. These roles describe how LM can regulate composition, carrier supply, interfacial coupling, and structural confinement.

The surrounding sectors classify the representative forms and dimensionalities included in the LMDS framework, including core-shell objects, porous or network structures, solution-derived forms, three-dimensional shapes, two-dimensional surfaces, one-dimensional wires, and zero-dimensional particles. This organization links LM morphology to possible structural, compositional, and interfacial degrees of freedom.

The bottom sequence highlights the processing-to-operation sequence. A room-temperature LM state provides processability and structural freedom; alloying, interfacial reconstruction, confinement, nanostructuring, or cooling then yields a transformed LM-derived structure; superconducting behavior is evaluated in the low-temperature state. The schematic panels at the right summarize generic superconducting signatures, including resistive transition, critical-field boundary, field-dependent resistance, and anisotropic upper critical field.

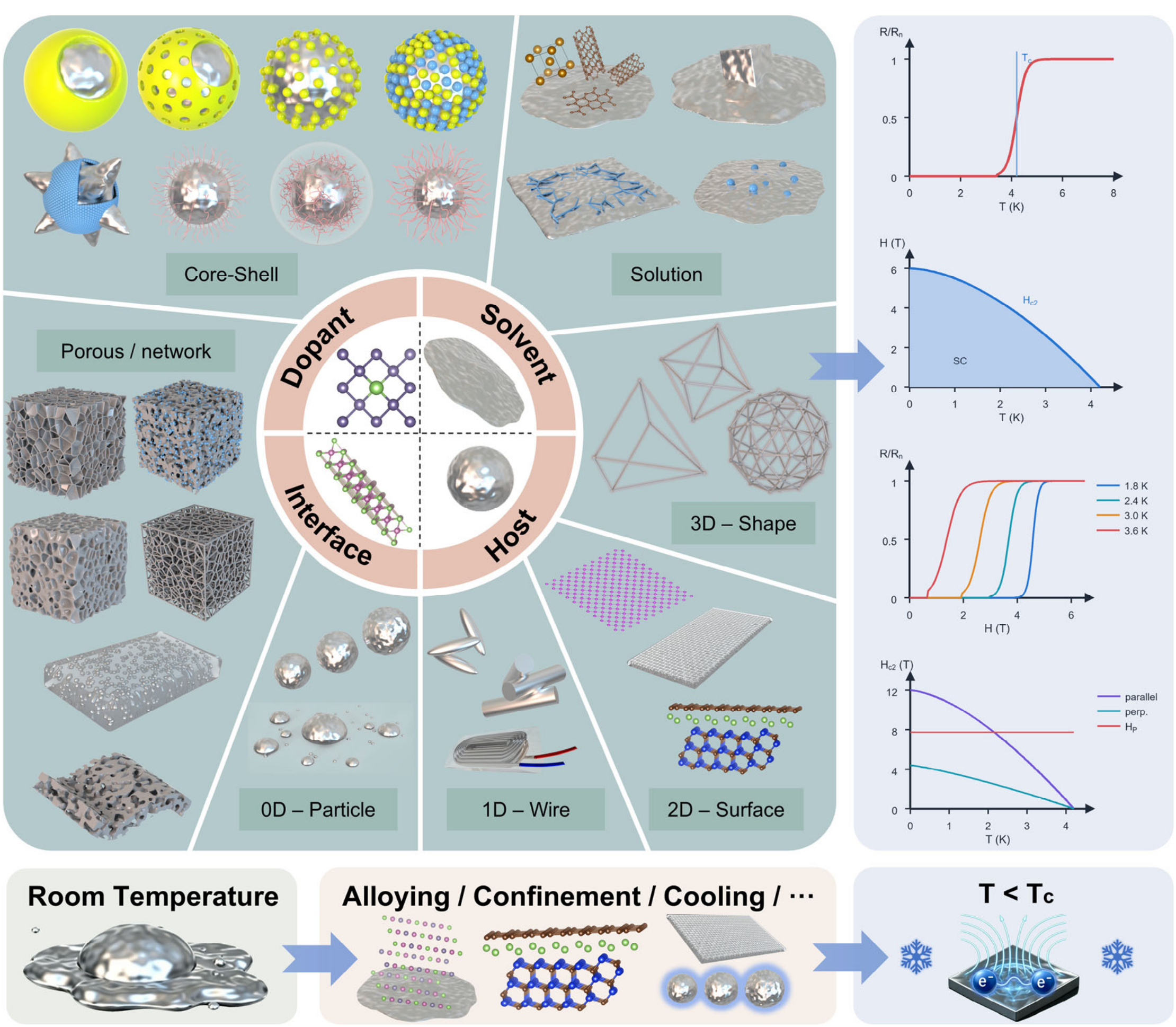

**Fig. 2 Schematic of mechanism-linked roles of LM in LMDS, showing four LM roles and their transformation into diverse architectures with superconducting functionality evaluated in the cooled state.**

The LM framework integrates synthesis, doping, interface formation, and intrinsic superconductivity within a single fluidic system, offering a basis for LMDS with diverse compositions and architectures. At this stage, the fluidity of LMs should be understood primarily as an advantage for room-temperature processing, patterning, and structural reconfiguration; the superconducting state generally emerges after cooling into solidified, amorphous, nanoconfined, or interfacially reconstructed phases. These features motivate the following discussion of fabrication strategies, device opportunities, disorder-enhanced superconductivity, and the open question of superconductivity in a true liquid state.

## 3 Paradigm for making LMDS

LMs are reshaping how superconducting-relevant material states can be fabricated by replacing high-temperature, instrument-intensive processing with liquid-state operations that are often performed at room temperature or under mild thermal conditions. Their low melting points, metallic bonding, and high interfacial mobility provide a dynamic reaction environment in which alloying, diffusion, intercalation, shaping, and nanodroplet formation can proceed with reduced thermal input. In this framework, LMs are described as active precursors, dopant reservoirs, interfacial reaction media, and structural templates for solidified, doped, nanoconfined, or interfacially reconstructed superconducting states.

**Fig. 3** organizes these routes in a four-column matrix: Column **i** shows the LM-enabled processing route, column **ii** specifies the processing condition, column **iii** identifies the LM-derived structure, and column **iv** presents the superconducting mechanism or experimental evidence. This row-column structure is important because it separates room-temperature processability from cryogenic superconducting verification.

In the bulk route (**Fig. 3a**), liquid-liquid interfacial reduction and diffusion at 25-80 °C under ambient pressure enable isothermal solidification and elemental redistribution in compositionally complex alloy particles[29]. Room-temperature, ambient-pressure shaping further illustrates how LM inks can form architected geometries without the thermal budgets required for conventional metal processing[30]. These operations establish alloy/intermetallic or shaped LM-derived structures; Superconducting behavior is represented separately by low-temperature resistance and ZFC/FC magnetization signatures from the superconducting LM alloys [27].

In the film route (**Fig. 3b**), Ga-source growth introduces substitutional Ga into Ge at approximately 100 °C under ultrahigh vacuum, producing Ga:Ge/Ge(001) films with carrier-density and lattice/phonon modification[22]. Room-temperature LM printing provides a complementary ambient route for defining metallic film or circuit geometries [27]. The superconducting mechanism in column **iv** is tied specifically to Ga-hyperdoped Ge, where Fermi-surface features and Raman phonon softening support carrier-mediated electron-phonon coupling[22].

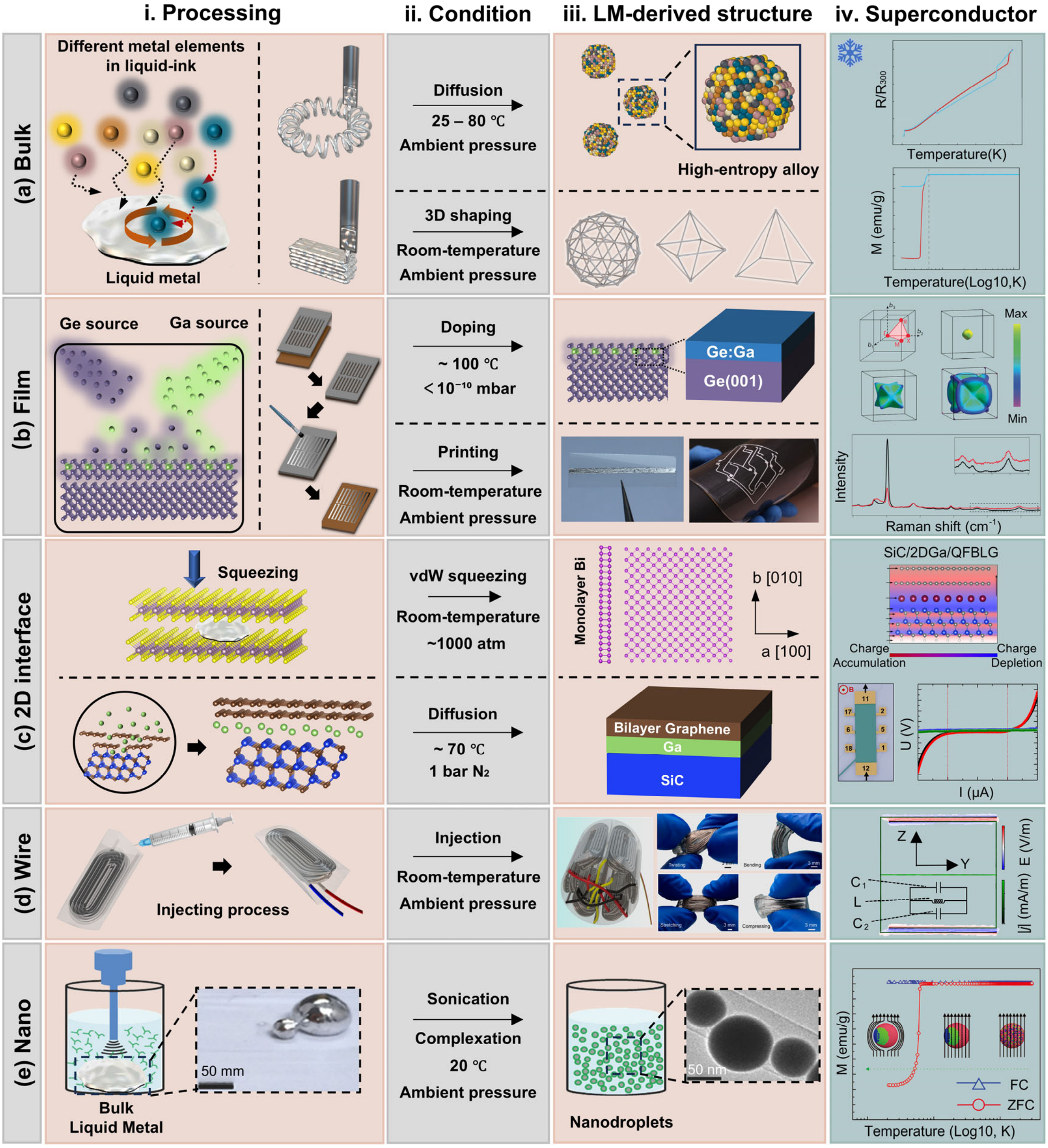

**Fig. 3 Processing-structure-superconductivity matrix for LMDS. Columns i-iv show processing, condition, LM-derived structure, and superconducting evidence/mechanism; rows (a)-(e) show bulk, film, two-dimensional interface, wire, and nano routes.** Fig. 3a(iv), Fig. 3b(i, right), and Fig. 3b(iii) are adapted from Ref. [27] with permission from IOP Publishing Ltd, Copyright 2025. Fig. 3b(iv) is adapted from Ref. [22] with permission from Springer Nature Limited, Copyright 2025. Fig. 3c(iv) is adapted from Ref. [23] with permission from Wiley-VCH GmbH, Copyright 2026. Fig. 3d(i) and Fig. 3d(iii) are adapted from Ref. [31] with permission from Springer Nature, Copyright 2024. Fig. 3e(i), Fig. 3e(iii), and Fig. 3e(iv) are adapted

from Ref. [17] with permission from Wiley, Copyright 2016. Fig. 3d(iv) is adapted from Ref. [28], licensed under a Creative Commons Attribution 4.0 International (CC BY 4.0) License.

In the two-dimensional interface route (**Fig. 3c**), room-temperature vdW squeezing under confinement pressure of approximately 1000 atm produces angstrom-thick two-dimensional metal sheets[32]. In a distinct intercalation pathway, Ga diffusion at about 70 °C under 1 bar $N_2$ forms Ga-intercalated epitaxial graphene devices[23]. The superconducting evidence in column **iv** is assigned to the graphene route, where calculated charge transfer and magnetotransport/critical-current signatures characterize superconductivity in the Ga-intercalated interface[23].

In the wire route (**Fig. 3d**), room-temperature injection under ambient pressure fills predesigned microchannels or capillaries to form continuous metallic pathway geometries[31]. This process demonstrates room-temperature geometric integration of liquid-metal pathways, whereas superconducting functionality must be established separately through cryogenic device measurements. In column **iv**, such functionality is represented by LM-derived resonator responses[28].

In the nano route (**Fig. 3e**), room-temperature sonication/complexation at approximately 20 °C under ambient pressure disperses EGaInSn/Ga-In-Sn into nanodroplets[17]. The resulting nanodroplet state is connected to superconductivity through low-temperature ZFC/FC magnetization signatures near $\boldsymbol{T_c}$ = 6.6 K[17].

Together, these approaches define a coherent LMDS paradigm in which room-temperature or near-room-temperature liquid-metal operations first establish composition, geometry, interfaces, or nanoscale states, while superconducting behavior is subsequently verified through low-temperature transport, magnetic, or resonator measurements. This processing-to-operation framework highlights a route toward efficient synthesis, free-form shaping, and integration of LMDS into reconfigurable superconducting and quantum systems.

Beyond these experimental advances, computational and data-driven methods are accelerating the discovery and optimization of LMDS.

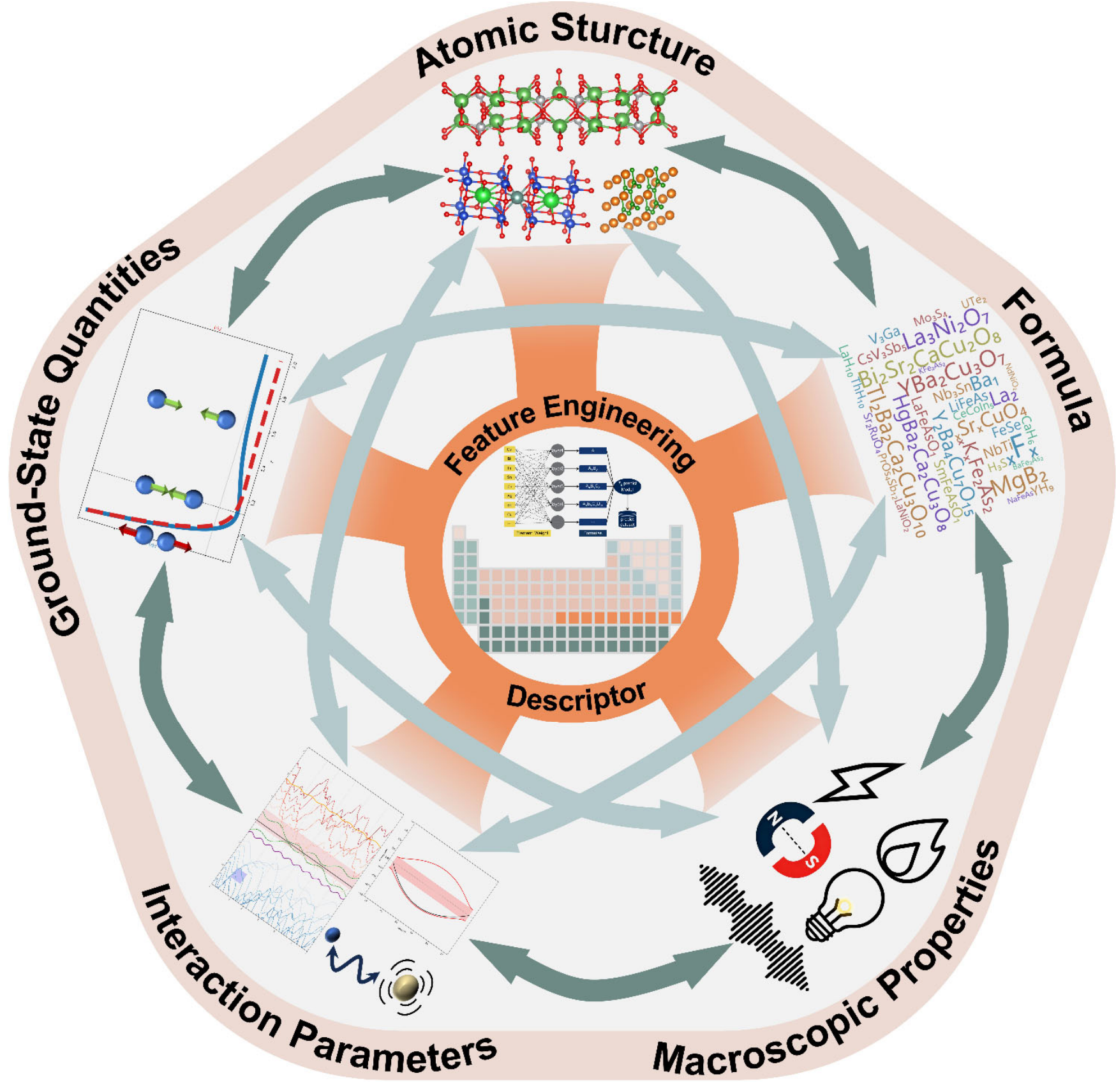

**Fig. 4. Pentagonal data framework enabling machine-learning-driven design and prediction of superconductors.**

Machine learning models trained on superconducting materials data are now identifying promising LMDS. This data-driven framework, conceptualized as part of the LM Materials Genome[33], integrates empirical knowledge with predictive computation. As illustrated in **Fig. 4**, five principal data domains underpin current superconductivity modeling: (i) Chemical composition, (ii) Atomic structure, (iii) Ground state quantities such as total energies, electronic bands and interatomic forces, (iv) Interaction parameters including superconducting gaps, electron–phonon coupling and phonon spectra, and (v) Macroscopic properties such as $\boldsymbol{T_c}$ and $\boldsymbol{T_m}$.

Crucially, the major domains within the proposed Materials Genome framework are already supported by curated and continuously expanding databases. Experimental repositories such as SuperCon$^2$[34], the MDR SuperCon Datasheet[35], and the HTS Material Properties Database[36] provide extensive records of chemical compositions, processing conditions, and measured superconducting properties. Structural and electronic information is available from databases including 3DSC[37], SuperBand[38], HTSC-2025[39], and the Kagome superconductor database[40], which contain crystal structures, electronic band structures, and Fermi-surface data.

Computational and interaction-related descriptors are further supported by resources such as the Conventional Superconducting Materials Database[41], JARVIS-BCS[42], JARVIS SuperconDB[43], BETE-NET[44], supercond-EPW[45], CriticalFieldDB[46], Superhydra Database[47], and the Heusler Superconducting Dataset[48], providing electron–phonon coupling parameters, Eliashberg spectral functions, and critical-field information. Recent studies on machine-learning-assisted superconductor discovery have also emphasized the importance of data standardization and physics-informed modeling for improving predictive reliability across diverse superconducting systems[49]. Collectively, these resources establish a substantial data foundation for the forward prediction and inverse design of LMDS.

These five domains are interconnected through bidirectional inference, enabling both forward prediction and inverse design. When the relationship between composition and structure is straightforward, $\boldsymbol{T_c}$ can be directly predicted by composition[50]; Conversely, inverse schemes can identify compositions yielding a targeted $\boldsymbol{T_c}$. In systems with multiple possible structures, structural prediction [51] followed by $\boldsymbol{T_c}$ estimation refines candidate selection. Ground-state quantities obtained from atomistic structure, such as a neuroevolution potential[52] trained on first principles calculation data, enhance the efficiency of molecular dynamics simulations and support reliable extraction of interaction parameters. Furthermore, electronic band features [38] and electron–phonon interactions serve as intermediate predictors of $\boldsymbol{T_c}$.

In particular, interaction parameters, such as $\lambda$ (the electron–phonon coupling constant), $\mu^*$ (the Coulomb pseudopotential), $\omega_{\log}$ (the logarithmic average phonon frequency), and $\overline{\omega}_2$ (the second-moment frequency), as well as their combinations, can be used to modify the equation(Eq. 6) proposed by McMillan[12] for calculating $\boldsymbol{T_c}$.

Xie et al.[53] introduced correction factors $f_\omega$ and $f_\mu$ through machine learning regression (Eq. 7-9), which inherit the key physical characteristics of the $f_1$ and $f_2$ correction parameters in the Allen–Dynes equation[54]. This approach significantly reduces the computational error of $\boldsymbol{T_c}$ while maintaining model simplicity and interpretability:

$$\mathrm{T}_c^{ML} = f_w f_\mu T_c^{McMillan} \tag{7}$$

$$f_\omega = 1.92\left(\frac{\lambda + \frac{\omega_{\log}}{\overline{\omega}_2} - \sqrt[3]{\mu^*}}{\sqrt{\lambda\, exp\left(\frac{\omega_{\log}}{\overline{\omega}_2}\right)}}\right) - 0.08 \tag{8}$$

$$f_\mu = \frac{6.86\exp\left(\frac{-\lambda}{\mu^*}\right)}{\frac{1}{\lambda} - \mu^* - \frac{\omega_{\log}}{\overline{\omega}_2}} + 1 \tag{9}$$

The integration of these components within the pentagonal data framework enables rapid and rational exploration of LMDS.

## 4 Future Perspectives

### *4.1 Reimagining Superconducting Devices with LMs*

LMSC introduce a new paradigm for superconducting device engineering by separating device fabrication from superconducting operation. During room-temperature fabrication and handling, the material exhibits liquid-like properties such as stretchability, self-healing, and conformal processability, enabling printing, casting, and reconfigurable patterning. Upon cooling to cryogenic temperatures, however, the LM solidifies into a rigid and structurally uniform superconducting phase. Thus, the soft mechanical properties associated with the liquid state and the zero-resistance superconducting state occur sequentially rather than concurrently. This distinction is essential for understanding the fabrication and operating characteristics of LM-based superconducting devices (Fig. 5).

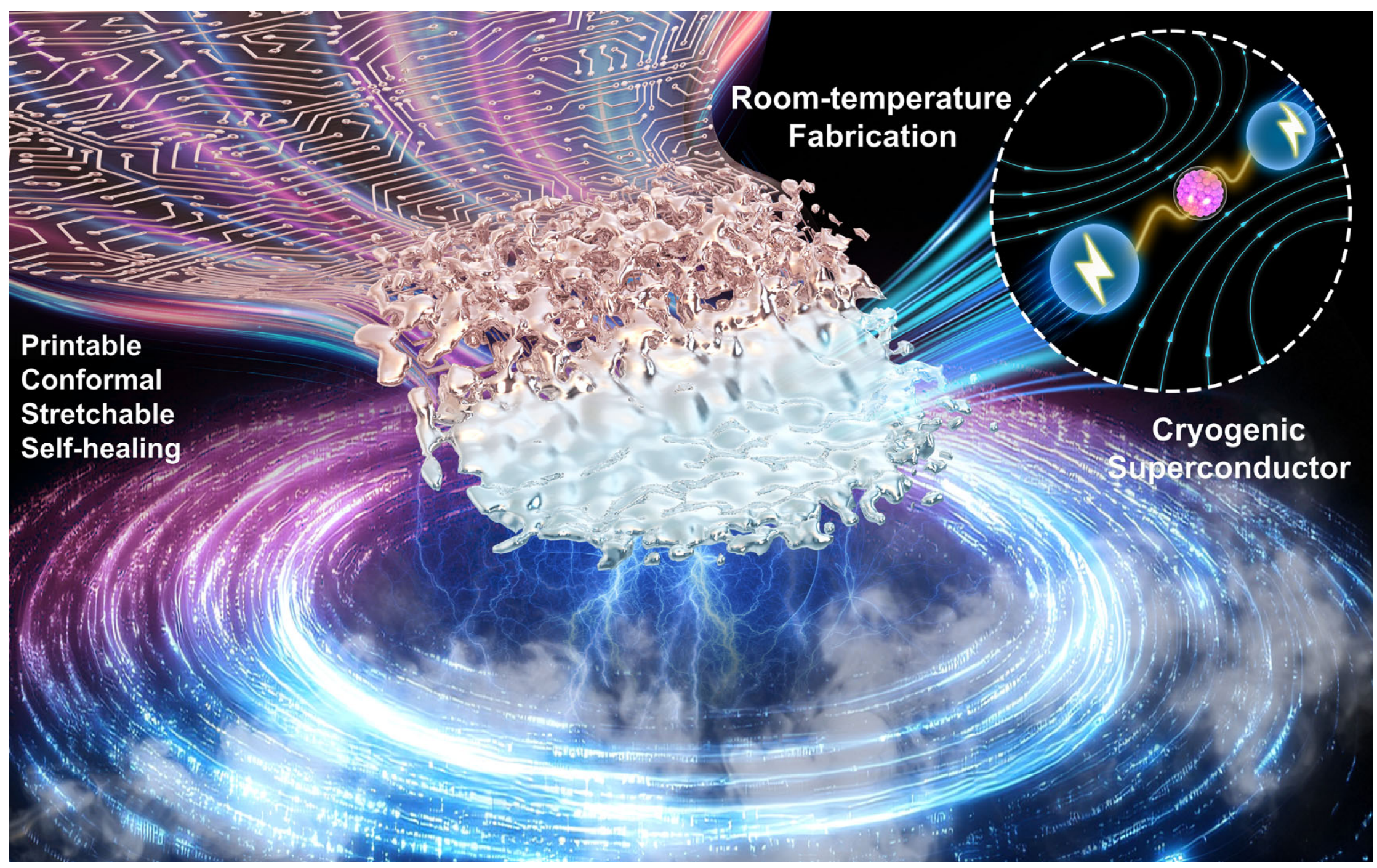

**Fig. 5 Dual-phase paradigm of LMDS: Liquid-state fabrication and cryogenic superconducting operation.**

In quantum and cryogenic applications, LM superconductors represent promising candidates for interconnects[27], coupling elements, resonators and Josephson junctions, where mechanical or thermal mismatch often limits reliability. Because their conductivity and electronic phase can be precisely tuned by temperature, the same material can reversibly switch between a superconducting state and a highly conductive normal state. The temperature-dependent transition between normal and superconducting states may facilitate device fabrication, assembly, testing, and operation within a unified material platform. Such thermal tunability may be advantageous for cryogenic systems that undergo repeated cooling and warming cycles.

Beyond the superconducting phase, LMs are well-established as flexible and deformable conductors for soft electronics near room temperature. Simple fabrication methods such as printing enable rapid production of LM wires, thin films, and circuits[55], laying the foundation for scalable manufacturing of superconducting LM devices. Additionally, LMs provide an efficient and low-cost strategy for manufacturing high-quality electronic devices[56,57]. The nanotip method can even enhance the printing precision of LMs down to 10 nanometers[58], enabling the fabrication of miniaturized devices.

The intrinsic fluidity of LMs at ambient conditions facilitates stretchable wiring, self-healing[59] of minor defects, and conformal interconnects on nonplanar substrates[60]. Integrating these characteristics with low-temperature superconductivity provides a route toward mechanically compliant and highly reliable re-configurational superconducting systems, allowing for the design of customized flexible sensor arrays or smart membrane layers for use in biomedical, aerospace, and other adaptive technologies.

Research has been conducted on flexible superconducting devices, including superconducting resonators[61,62], and superconducting quantum interference devices (SQUIDs)[63]. The response speed and quantum coherence of such devices are highly dependent on material dimensions, making miniaturization a key driving direction in the field[64,65]. However, existing flexible devices are fabricated by depositing rigid superconductors onto flexible substrates, which does not fundamentally transcend the paradigm of rigid superconductor processing and entails significant complexity in the fabrication of microscale circuits. The precisely printable and flexible properties of LMs near room temperature bring new opportunities to the field of flexible superconductivity, offering novel fully flexible fabrication approaches for such superconducting devices, and even for complex integrated quantum circuits, including single-photon detectors and superconducting qubit chips.

The reconfigurable nature of soft, transformable LMs[66], combined with their stable electrical performance near room temperature, allows reversible and programmable control has achieved advances in sensing applications including antennas. In quantum computing, ambient-temperature fluidic manipulation enables dynamic adjustment of qubit connectivity, optimizing the match between hardware and algorithms. Cooling to the superconducting state, the system locks into an optimal configuration for executing highly complex simulations or machine learning tasks, significantly enhancing efficiency and robustness.

Overall, LMDS provide a new framework for superconducting device engineering in which fabrication and operation are separated into liquid-state and superconducting phases, respectively. By combining ambient-temperature processability with cryogenic functionality within a single material system, LMDS may enable device architectures and manufacturing strategies that are difficult to realize using conventional superconducting materials.

### *4.2 Disorder-Enhanced Superconductivity in LMDS*

Most established superconductors have traditionally been developed within a crystalline-structure paradigm, where long-range atomic order is used to control electronic structure, phonon behavior, carrier concentration, and pairing interactions. In conventional superconductors such as A15 compounds and $MgB_2$-type systems, ordered crystal structures support characteristic electronic and phononic features associated with superconductivity. Even in unconventional superconductors such as cuprates and Fe-based systems, specific structural motifs remain closely related to the underlying electronic structure and competing electronic orders.

Within this framework, structural disorder is often viewed as detrimental because defects and random scattering can broaden electronic states, reduce carrier mobility, suppress coherence, and in some cases act as pair-breaking perturbations. Excessive disorder may also lead to localization, granularity, or filamentary transport behavior.

LMDS provide an opportunity to extend this conventional picture. The amorphous Ga and Bi examples discussed in **Section 2** suggest that superconductivity can persist or even be enhanced in structurally disordered or metastable metallic phases. Because liquid metals possess low melting points, high atomic mobility, and accessible solidification pathways, their local structure can be tuned through cooling rate, composition, confinement, substrate interaction, and interfacial reconstruction. Rather than relying only on long-range periodicity, LMDS may also exploit short-range order, metastability, and local coordination environments.

Rapid quenching, amorphization, or nanoscale confinement may introduce local structural and vibrational modifications, including changes in coordination environment, bond distribution, phonon spectra, and electronic density of states, which can influence electron–phonon coupling and superconducting behavior. From this perspective, disorder in LMDS is not simply a defect source, but a potentially tunable structural variable linked to processing history and local bonding environments.

This disorder-assisted strategy may complement conventional crystalline superconducting design. Relevant variables include not only composition and average structure, but also quenching history, glass-forming tendency, nanoscale confinement, interfacial bonding, and structural relaxation. At the same time, maintaining phase coherence and distinguishing genuine superconductivity from filamentary transport remain important experimental challenges.

### *4.3 Open Question: True Liquid Superconductors*

Within the LMDS framework developed in this perspective, LMs function as reaction media, dopant sources, interfacial mediators, superconducting hosts, and data-rich materials spaces for predictive design. Their amorphous, alloyed, interfacial, nanoconfined, and otherwise transformed superconducting states raise a sharper question: What kind of pairing mediator could support superconductivity in a thermodynamically homogeneous liquid. Earlier discussions treated liquid superconductivity as exceptional and unresolved[67]. Related proposals likewise emphasize its unusual requirements[68]. To make this question physically tractable, true liquid superconductivity should first be separated from the static-disorder limits already represented by amorphous and dirty superconductors.

The first distinction is temporal disorder. Amorphous and dirty superconductors can tolerate static disorder when the ionic configuration is frozen on the pairing time scale[69]; The amorphous Ga and Bi systems[13,14] discussed here belong to this reference class. A liquid is different because diffusion, cage breaking, and structural relaxation make the ionic potential time dependent. The issue is whether moving disorder can retain a retarded attractive channel before it becomes pair-breaking noise.

A true-liquid theory must start from a dynamic spectrum. The proposed LM materials genome should retain composition, short-range order, electronic structure, interaction parameters, and macroscopic $\boldsymbol{T_c}$ or $\boldsymbol{T_m}$, but add $S(q,\omega)$, current correlations, tau_alpha, and linewidths. Liquid-metal spectroscopy shows that collective modes and relaxation channels can coexist in the same spectrum[70]. Liquid Ga provides a concrete case: Acoustic-like modes are present, but their damping is strongly wave-vector dependent[71]. A first criterion is local in frequency and momentum: Spectral weight should satisfy $\Omega(q)\tau_\alpha > 1$ and $\Gamma(q) < \Omega(q)$ on the pairing time scale. Here $\tau_\alpha$ connects pairing to structural relaxation rather than to a frozen lattice[72].

A phonon-like mechanism is the most conservative route, but it requires particular caution. It must generate an attractive kernel that remains retarded after Coulomb screening and underdamped in the relevant frequency window. Liquid phonon descriptions support only limited windows of phonon-like behavior[73], while recent liquid theories emphasize that low-frequency shear response may be overdamped[74]. Metallic-liquid $T_c$ calculations are therefore best viewed as formal guidance: They use liquid density correlations or effective kernels, not an imported crystal phonon spectrum[75]. The same density motion must also be tested against pair breaking, since scattering can suppress the condensate[76].

Non-phonon mechanisms widen the search space but do not relax the standard. Spin or quantum-critical fluctuations can mediate pairing in unconventional solids[77], and spin-fluctuation theory shows how a repulsive interaction can produce structured pairing[78]. Plasmonic routes indicate that charge fluctuations may also be relevant [79]. In a liquid, each proposal must specify the fluctuation spectrum, the screened pairing vertex, the competing inelastic scattering rate, and the expected phase stiffness. Incoherent-pairing theories show that pairing without long-lived quasiparticles can occur in some limits[80], while pair-breaking dynamics remains a central constraint [81].

The final filter is phase coherence. A local pair, transient gap, or pseudogap-like response is not macroscopic superconductivity unless long-range phase rigidity develops[82]. Recent phase-fluctuation frameworks reinforce this distinction between local pairing signatures and coherent order[83]. Accordingly, LM-derived systems should be interpreted as reference platforms: Nanoconfined Ga probes confinement and metastability[84], Ga-based nanodroplets probe deformable superconducting architectures[17], and Ga-In-Sn alloys probe low-melting alloy behavior[19]. They do not establish a homogeneous liquid condensate.

This framing turns speculation into variables. Pressure, confinement, alloying, and supercooling can tune diffusivity, tau_alpha, short-range order, screening length, and the damping structure of $S(q, \omega)$. Metallic hydrogen offers only a distant analogy: Quantum-fluid pairing under compression[85], possible superconductor-superfluid physics[86], early liquid-metallic-hydrogen modeling[87], and recent high-pressure estimates[88]. These analogies are not templates for LMs. A future true-liquid claim would have to close the loop in one volume and one thermodynamic state: In situ structural evidence of liquidity, four-probe zero resistance, bulk magnetic shielding or a Meissner response, a thermodynamic anomaly at $\boldsymbol{T_c}$, and exclusion of crystallites, oxide skins, frozen layers, inclusions, filaments, and percolative paths[89].

## 5 Conclusion

LMs provide a versatile and conceptually very promising platform for realizing superconductivity. In this perspective, we have shown that a single LM can act as solvent, dopant source, interfacial mediator and superconducting host, and we organized these roles into a unified LM paradigm for fabricating superconductors under near-ambient conditions. Within this framework, we outlined the concrete LM routes for obtaining bulk alloys, thin films, two-dimensional confined phases, wires, nanodroplets and printed circuits. These LMDS can be formed with low energy input while exhibiting flexibility, self-healing and conformal contact that are difficult to achieve with brittle ceramic or high-melting metallic superconductors.

We further introduced an AI-assisted LM materials genome framework that links composition, structure, ground-state quantities, interaction parameters and macroscopic properties such as $\boldsymbol{T_c}$ and $\boldsymbol{T_m}$. This pentagonal data framework provides a rational path to both forward prediction and inverse design of LMDS and offers a systematic strategy that moves beyond empirical trial-and-error synthesis.

From a fundamental viewpoint, LM systems, especially gallium-based amorphous and nanoconfined states, supply a controllable environment for exploring superconductivity in the presence of strong disorder, metastability and structural dynamism. These systems help clarify how far superconducting coherence can persist without long-range crystalline order and bring the open question of superconductivity in a true liquid state into a concrete experimental context. The possibility of stabilizing superconducting behavior in overpressurized or nanoconfined LMs provides a valuable direction for investigating pairing mechanisms in dynamically disordered phases.

Overall, this perspective positions LMs not only as convenient processing media but as a coherent materials and design space for flexible, energy-efficient and reconfigurable superconducting technologies. By combining the LM paradigm with data-driven discovery and targeted experiments, a systematic route emerges for developing new superconducting compositions and architectures and for deepening our understanding of superconductivity in amorphous, disordered and potentially liquid-like states.

## Acknowledgments

Not Applicable

## Disclosure statement

No potential conflict of interest was reported by the author(s).

## Data availability

Not Applicable

## Author contributions

Chen Hua: Conceptualization (Equal); Data curation (Lead); Formal analysis (Lead); Investigation (Lead); Methodology (Lead); Validation (Lead); Visualization (Lead); Writing – original draft (Lead); Writing – review and editing (Supporting).

Wendi Bao: Investigation (Supporting); Visualization (Supporting); Writing – original draft (Supporting); Writing – review and editing (Supporting).

Minghui Guo: Visualization (Supporting); Writing – review and editing (Supporting).

Jing Liu: Conceptualization (Equal); Project administration (Lead); Supervision (Lead); Visualization (Supporting); Writing – original draft (Supporting); Writing – review and editing (Supporting).